\documentclass[final,onecolumn]{IEEEtran}

\usepackage{fancyhdr}
\usepackage{latexsym}
\usepackage{graphicx,color}
\usepackage{colortbl,latexsym}
\usepackage{epstopdf}  

\usepackage{shadow,epsf,amsthm,amssymb,amsmath}

\usepackage{enumerate}

\usepackage{setspace}                           

\usepackage{lipsum}

\usepackage{tikz}	
\usetikzlibrary{backgrounds,fit,decorations.pathreplacing,chains}  

\usepackage[colorinlistoftodos]{todonotes}

\newtheorem{definitionenv}{Definition}
\newtheorem{lemmaenv}[definitionenv]{Lemma}
\newtheorem{theoremenv}[definitionenv]{Theorem}
\newtheorem{corollaryenv}[definitionenv]{Corollary}
\newtheorem{propositionenv}[definitionenv]{Proposition}
\newtheorem{conjectureenv}[definitionenv]{Conjecture}
\newtheorem{remarkenv}[definitionenv]{Remark}
\newenvironment{remark}{\begin{remarkenv}\rm}{\end{remarkenv}}
\newcommand{\br}{\begin{remark}}
\newcommand{\er}{\end{remark}}

\newtheorem{exampleenv}{Example}
\newtheorem{app-lemmaenv}[section]{Lemma}

\newenvironment{definition}{\begin{definitionenv}\rm}{\end{definitionenv}}
\newenvironment{lemma}{\begin{lemmaenv}\rm}{\end{lemmaenv}}
\newenvironment{theorem}{\begin{theoremenv}\rm}{\end{theoremenv}}
\newenvironment{corollary}{\begin{corollaryenv}\rm}{\end{corollaryenv}}
\newenvironment{example}{\begin{exampleenv}\rm}{\end{exampleenv}}
\newenvironment{proposition}{\begin{propositionenv}\rm}{\end{propositionenv}}
\newenvironment{conjecture}{\begin{conjectureenv}\rm}{\end{conjectureenv}}
\newenvironment{app-lemma}{\begin{app-lemmaenv}\rm}{\end{app-lemmaenv}}

\newcommand{\bd}{\begin{definition}}
\newcommand{\ed}{\end{definition}}
\newcommand{\bl}{\begin{lemma}}
\newcommand{\el}{\end{lemma}}
\newcommand{\elp}{\hspace*{\fill} $\Box$
                 \end{lemma}}
\newcommand{\bt}{\begin{theorem}}
\newcommand{\et}{\end{theorem}}
\newcommand{\etp}{\hspace*{\fill} $\Box$
                 \end{theorem}}
\newcommand{\bc}{\begin{corollary}}
\newcommand{\ec}{\end{corollary}}
\newcommand{\ecp}{\hspace*{\fill} $\Box$
                 \end{corollary}}
\newcommand{\bcj}{\begin{conjecture}}
\newcommand{\ecj}{\end{conjecture}}

\newcommand{\be}{\begin{example}}
\newcommand{\ee}{\end{example}}
\newcommand{\eep}{\hspace*{\fill} $\Box$
                 \end{example}}
\newcommand{\bp}{\begin{proposition}}
\newcommand{\ep}{\end{proposition}}
\newcommand{\epp}{
                 \end{proposition}}

\newcommand{\bra}[1]{\langle#1|}
\newcommand{\ket}[1]{|#1\rangle}
\newcommand{\braket}[2]{\langle#1|#2\rangle}

\newcommand{\tr}[1]{\text{tr}\left(#1\right)}

\newcommand{\diag}[1]{\text{diag}\left(#1\right)}
\newcommand{\rank}[1]{\text{rank}\left(#1\right)}

\newcommand{\eeq}{ \setcounter{equation} {\value{enumi}}}

\newcommand{\cA}{\mathcal{A}}
\newcommand{\cB}{\mathcal{B}}
\newcommand{\cC}{\mathcal{C}}

\newcommand{\cH}{\mathcal{H}}
\newcommand{\cI}{\mathcal{I}}
\newcommand{\cL}{\mathcal{L}}
\newcommand{\cN}{\mathcal{N}}
\newcommand{\cM}{\mathcal{M}}

\newcommand{\cQ}{\mathcal{Q}}
\newcommand{\cR}{\mathcal{R}}

\newcommand{\Tr}[1]{\text{Tr}\left(#1\right)}
\def\beq{\begin{equation}}
\def\eeq{\end{equation}}

\def\bean{\begin{IEEEeqnarray*}{rCl}}
\def\eean{\end{IEEEeqnarray*}}

\begin{document}


\title{On the One-Shot Zero-Error Classical Capacity of Classical-Quantum Channels Assisted by Quantum Non-signalling Correlations}

\author{Ching-Yi Lai and Runyao Duan
\thanks{
The authors are with the Centre for Quantum Computation \& Intelligent Systems, Faculty of Engineering and Information Technology, University of Technology, Sydney, NSW, 2007, Australia.
Runyao Duan is also with the UTS-AMSS Joint Research Laboratory of Quantum Computation and Quantum Information Processing,
Academy of Mathematics and Systems Science, Chinese Academy of Science, Beijing 100190, China.
(emails: cylai0616@gmail.com and Runyao.Duan@uts.edu.au)
}
}

%

\date{\today}

\maketitle

\begin{abstract}
Duan and Winter studied the one-shot zero-error classical capacity of  a quantum channel assisted by quantum non-signalling correlations,
and formulated this problem as a semidefinite program depending only on the Kraus operator space of the channel.
For the class of classical-quantum channels, they showed that the \emph{asymptotic} zero-error classical capacity assisted by quantum non-signalling correlations, 
 minimized over all classical-quantum channels with a confusability graph $G$,
is exactly $\log \vartheta(G)$, where $\vartheta(G)$ is the celebrated  Lov\'{a}sz theta function.
\textcolor[rgb]{0.00,0.00,0.00}{In this paper, we show that   the \emph{one-shot} capacity for a  classical-quantum channel, induced from a \emph{circulant graph} $G$ defined by \emph{equal-sized cyclotomic cosets}, is $\log \lfloor \vartheta(G) \rfloor$,
which further implies that its \emph{asymptotic} capacity  is $\log \vartheta(G)$.}
This type of graphs include the cycle graphs of odd length, the Paley graphs of prime vertices, and the cubit residue graphs of prime vertices.
Examples of other graphs are also discussed.
This endows the Lov\'{a}sz $\theta$ function with a more straightforward  operational meaning.
\end{abstract}

\section{Introduction}
Shannon discussed the communication problem in the setting of zero errors
and  connected this problem to the graph theory~\cite{Shannon56}.
Let $N:V\rightarrow W$ be a channel with discrete alphabets $V$ and $W$.
We want to determine the maximum messages that can be sent through the channel $N$ without confusion.
Two distinct messages can be confused if their channel outputs are equal with a nonzero probability.
It turns out that the maximum distinguishable messages is equal to the largest number of independent vertices $\alpha(G)$ of its \emph{confusability  graph} $G$.
The confusability graph $G$ of channel $N$ has a vertex set $V$, which is the channel input alphabet,
and an edge set $E$ so that two vertices $v$ and $w$ are connected (say, $vw\in E$) if their channel outputs are likely to be confused.
Using the channel $N$ twice in parallel corresponds to a confusability graph $G\boxtimes G$, where
$\boxtimes$ is the graph \emph{strong product}.
(For two graphs $G_1, G_2$ with vertex sets $V_1, V_2$, and edge sets $E_1,E_2$, respectively, their {strong product}  $G_1\boxtimes G_2$ has
a vertex set $V_1\times V_2$, and two vertices $(v_1,v_2)$ and $(w_1,w_2)\in V_1\times V_2$ are connected if $v_1w_1\in E_1$ and  $v_2w_2\in E_2$;   or  $v_1w_1\in E_1$  and $v_2=w_2$;   or
$v_1=w_1$ and  $v_2w_2\in E_2$.) 
The \emph{Shannon capacity} of a graph $G$ is defined as
\begin{align}
\Theta(G)=\sup_n \sqrt[n]{\alpha(G^{\boxtimes n})}= \lim_{n\rightarrow \infty} \sqrt[n]{\alpha(G^{\boxtimes n})}.
\end{align}
The quantity $\Theta(G)$ is difficult to determine, even for simple graphs, such as cycle graphs $\cC_n$ of odd length.
In \cite{Lov79}, Lov\'{a}sz proposed an upper bound $\vartheta(G)$ (to be defined in Sec. \ref{sec:Lov}) on $\Theta(G)$, and it is tight in some cases.
For example, $\Theta(\cC_5)=\vartheta(\cC_5)$. Although $\Theta(\cC_n)$ for odd $n\geq 7$ are still unknown, it seems close to $\vartheta(\cC_n)$.
However, Haemers showed that it is possible that there is a gap between $\vartheta(G)$ and $\Theta(G)$ for some graphs \cite{Haemers79,Haemers78}.
It is desired to find operational meanings for  $\vartheta(G)$, apart from an upper bound for $\Theta(G)$.

Recently the problem of zero-error communication  has been studied in   quantum information theory \cite{MACA06,DSW13}.
Some unexpected phenomena were observed in the quantum case. For example, very noisy channels can be super-activated \cite{DS08,Duan09,CCH11,CS12}.
It is also likely that entanglement can increase the zero-error capacity of classical channels \cite{CLMW10,LMMOR12}.
Again, entanglement-assisted zero-error capacity is upper-bounded by the Lov\'{a}sz $\vartheta$ function \cite{Beigi10}.
For a classical channel, it is suspected that its entanglement-assisted zero-error capacity is exactly the  Lov\'{a}sz $\vartheta$ function \cite{DSW13}.

Non-signalling correlations have been studied in relativistic causality of quantum operations \cite{BGNP01,ESW02,PHHH06,OCB12,Chi12}.
In \cite{CLMW11}, Cubitt \emph{et al.} considered non-signalling correlations in the zero-error classical communications.
Duan and Winter further introduced quantum non-signalling correlations (QNSCs) in the zero-error communication problem \cite{DW14}.
QNSCs are completely positive and trace-preserving linear maps   shared between two parties 
so that they cannot send any information to each other  by using these linear maps.
Resources, such as shared randomness, entanglement, and classical non-signalling correlations, can be considered as special types of QNSCs.
The one-shot zero-error classical capacity of a quantum channel $\cN$ assisted by a QNSC $\Pi$
is  the \emph{logarithm} of the largest integer $m$ so that a noiseless classical channel that can send $m$ messages
can be simulated by the composition of $\cN$ and $\Pi$.
Duan and Winter formulated this problem as a semidefinite program (SDP) \cite{VB96}.
For the class of  classical-quantum (CQ) channels,
 the \emph{one-shot} zero-error classical capacity assisted by QNSCs is $\log \lfloor \Upsilon(\cN)\rfloor$, where $\Upsilon(\cN)$ is the value of an SDP (see Eq. (\ref{eq:SDP}) below) \cite{DW14}.
Moreover, they proved that the \emph{asymptotic} zero-error classical capacity assisted by QNSCs, minimized over all CQ channels with a confusability graph $G$,  is exactly $\log \vartheta(G)$.
This provides an operational meaning of the Lov\'{a}sz $\vartheta$ function.
(The definition of a confusability graph can be generalized to quantum channels. For CQ channels, see Sec. \ref{sec:zero_error_QNS}.)
In \cite{DW15}, they showed that $ \vartheta(G)$ is also the one-shot QNSC-assisted zero error capacity activated by additional forward noiseless classical channels,
minimized over all CQ channels with a confusability graph $G$.

In this article we focus on the same problem in the \emph{one-shot} setting.
We consider the type  of CQ channel
$\cN: \ket{k}\bra{k}\mapsto \ket{u_k}\bra{u_k}$, 
where $\{ \ket{u_k}\}$
is an \emph{orthonormal representation}  of a graph $G$  in some Hilbert space $\cB$.
We will provide a class of \emph{circulant graphs}, defined by \emph{equal-sized cyclotomic cosets}, and their orthonormal represntations so that
the one-shot QNSC-assisted zero-error classical capacity of a  CQ channel $\cN$ induced from these orthonormal representations is
$$\log \lfloor\Upsilon(\cN)\rfloor=\log \lfloor\vartheta(G)\rfloor. $$
Moreover, the asymptotic QNSC-assisted zero-error classical capacity of $\cN$ is
$$
C_{0,\text{NS}}(\cN)=\lim_{m\rightarrow \infty}\frac{1}{m}\log \Upsilon(\cN^{\otimes m})=\log \vartheta(G),
$$
since  $\Upsilon$ is super-multiplicative and $C_{0,\text{NS}}(\cN)$ is upper bounded by $\log \vartheta(G)$ (see Eq. (\ref{eq:C0N upper bound})).
This provides a more straightforward operational meaning for the Lov\'{a}sz $\vartheta$ function.
In particular, our results apply to the cycles $\cC_n$ of odd length.
There are some works trying to connect the Shannon capacity $\Theta(\cC_n)$ and independence number $\alpha(\cC_n^{\boxtimes m})$ to  $\vartheta(\cC_n)$   \cite{VZ02,CGR03,Boh03,Boh05}.
Now we know that with the assistance of quantum non-signalling correlations,  $\Upsilon(\cN)=\vartheta(\cC_n)$.
This may explain why it is difficult to build equality between $\Theta(\cC_n)$ and $\log \vartheta(\cC_n)$.

This paper is organized as follows. We first  give definitions of graphs, orthonormal representations, and the Lov\'{a}sz $\vartheta$ function in the next section.
QNSC-assisted zero-error communication  is introduced in Sec. \ref{sec:zero_error_QNS}.
In Sec. \ref{sec:circulant_graph}, we provide an orthonormal representation for any circulant graph. 
Then we explicitly construct an optimal feasible solution to the SDP for the one-shot QNSC-assisted zero-error classical capacity of a CQ channel, whose confusability graph
is a circulant graph defined by equal-sized cyclotomic cosets.
These circulant graphs are characterized in Sec. \ref{sec:characterization},
and they include three families of graphs:  the cycle graphs $\cC_n$ of odd length,
the Paley graphs $\cQ\cR_p$, where $p$ is a prime congruent to $1$ modulo 4,
and the cubic residue graphs $\cC\cR_p$, where $p$ is a prime congruent to $1$ modulo 3.
Finally we conclude with a discussion on other graphs with $\Upsilon(\cN)=\vartheta(G)$ in Sec. \ref{sec:discussion}.
\section{Lov\'{a}sz $\vartheta$ function and Graphs}
\label{sec:Lov}

In this article the vertex set $V$ of a graph $G$ under consideration is the ring of integers modulo $n$ for $n=|V|$. That is, $V=\mathbb{Z}/n\mathbb{Z}=\mathbb{Z}_n=\{0,1,\dots, n-1\}$.
Let $E$ be the edge set of $G$ and  let $vw$ denote an edge connecting vertices $v$ with $w$.
Let 
$[M]_{i,j}$ denote the $(i,j)$ entry of a matrix $M$.
The adjacency matrix $A_G$ of $G$ has entries \[
[A_G]_{i,j}=\left\{
                                                                 \begin{array}{ll}
                                                                   1, & \hbox{ if $ij\in E$;} \\
                                                                   0, & \hbox{otherwise.}
                                                                 \end{array}
                                                               \right.
\]
The eigenvalues and eigenvectors of a graph $G$ are the eigenvalues and eigenvectors of its adjacency matrix $A_G$.
An \emph{automorphism} on a graph $G$ is a permutation on its vertex set $V$ that preserves the adjacency.
Consequently, the adjacency matrix $A_G$ is invariant under the conjugation of an  automorphism.
A graph is called \emph{asymmetric} if it has no nonidentity automorphism.
If for any two edges of $G$, there exists an automorphism mapping one edge to the other, then $G$ is \emph{edge-transitive}.

In order to estimate $\Theta(G)$, Lov\'{a}sz proposed an upper bound $\vartheta(G)$ on the Shannon capacity of a graph $G$  \cite{Lov79},
which is the minimum \emph{value} of an \emph{orthonormal representation}  of the graph.
We use a more general definition of an orthonormal representation as follows.
\bd \label{def:OR2}
Suppose $\{P_k\} \in \mathbb{C}^{d\times d}$  is a set of $n$ orthogonal projectors so that
\[
\Tr{P_iP_j} =0
\]
if   $ij\notin E$. Then  $\{P_k\}$ is an orthonormal representation of $G$.
The {value} of $\{P_k\}$ is defined as
\[
\eta(\{P_k\})= \min_{{\sigma\geq 0: }\atop{\Tr{\sigma}=1}} \max_{k} \frac{1}{\Tr{P_k \sigma}}.
\]
(This definition of $\eta$ is different from that in \cite{DW14}.)
The trace-one, positive semidefinite operator $\sigma \in \mathbb{C}^{d\times d}$ that yields the minimum value is called the \emph{handle} of the representation.
Then $\vartheta(G)$  is defined as \[
\vartheta(G)= \min_{\{P_k\}}\eta(\{P_k\}).
\]
\ed
\noindent  We also say that $\vartheta(G)$ is the \emph{Lov\'{a}sz number} of $G$.
 An \emph{optimal} orthonormal representation (OOR) of $G$ is a representation with value $\vartheta(G)$.
If $P_k$ and $\sigma $ are restricted to rank-one matrices, this is exactly the definition in \cite{Lov79}.
When $P_k=\ket{u_k}\bra{u_k}$ and $\sigma=\ket{c}\bra{c}$, we also say that $\{\ket{u_k}\}$ is an orthonormal representation of $G$ with handle $\ket{c}$, without ambiguity.
Following \cite{DW14,CSW14}, one can show that the definition is well-defined even allowing $P_k$ and $\sigma$ to have rank greater than one.

\textcolor[rgb]{0.00,0.00,0.00}{In \cite{Lov79}, it is shown that $\alpha(G)\leq \vartheta(G)$. Furthermore, $\vartheta(G)$ is multiplicative:
\begin{align}
\vartheta(G\boxtimes H)=\vartheta(G)\vartheta(H) \label{eq:theta multiplicative}
\end{align}
for two graphs $G$ and $H$.  Therefore, it is easy to see that $\Theta(G)\leq  \vartheta(G)$.}

Finally, in  \cite[Theorem 3]{Lov79}, Lov\'{a}sz showed that  $\vartheta(G)$ is the minimum of the largest eigenvalue of any symmetric matrix $A$
such that \begin{align}
\mbox{$[A]_{i,j}=1$ if $i=j$ or $ij\notin E$.} \label{eq:lovasz_matrix}
\end{align}
Thus $\vartheta(G)$ can be determined by solving an SDP,
and it serves as a practical upper bound on $\Theta(G)$.

\section{Zero-Error Communication Assisted with Quantum Non-Signalling Correlations}
\label{sec:zero_error_QNS}
Let $\cL(\cH)$ denote the space of linear operators on Hilbert space $\cH$.
Quantum non-signalling correlations are completely positive and trace-preserving linear maps $\Pi: \cL(\cA_i)\otimes \cL(\cB_i)\rightarrow \cL(\cA_o)\otimes \cL(\cB_o)$
shared between two parties Alice and Bob (with Hilbert spaces $\cA$ and $\cB$, respectively, and the subscripts $i$ and $o$ stand for input and output, repectively) so that they cannot send classical information to each other  by using $\Pi$.
Let the Choi matrix of $\Pi$ be
\[
\Omega_{\cA_i'\cA_o\cB_i'\cB_o}= (\text{id}_{\cA_i'}\otimes \text{id}_{\cB_i'}\otimes \Pi) (\Phi_{\cA_i\cA_i'}\otimes \Phi_{\cB_i\cB_i'}),
\]
where $\text{id}_{\cA}\in\cL(\cA)$ denotes the identity operator on the Hilbert space $\cA$, $\Phi_{\cA_i\cA_i'}= \ket{\Phi_{\cA_i\cA_i'}}\bra{\Phi_{\cA_i\cA_i'}}$,
$\Phi_{\cB_i\cB_i'}= \ket{\Phi_{\cB_i\cB_i'}}\bra{\Phi_{\cB_i\cB_i'}}$,and $\ket{\Phi_{\cA_i\cA_i'}}=\sum_k \ket{k_{\cA_i}}\ket{k_{\cA_i'}}$ and
 $\ket{\Phi_{\cB_i\cB_i'}}=\sum_k \ket{k_{\cB_i}}\ket{k_{\cB_i'}}$ are the un-normalized  maximally-entangled states.
For $\Pi$ to be a QNSC, Duan and Winter derived the following constraints \cite{DW14}:
\begin{align*}
\Omega_{\cA_i'\cA_o\cB_i'\cB_o}\geq 0,&\\
\text{Tr}_{\cA_o\cB_o}\left({\Omega_{\cA_i'\cA_o\cB_i'\cB_o}}\right)=\mathbb{I}_{\cA_i'\cB_i'},&\\
\text{Tr}_{\cA_o\cA_i'}\left({\Omega_{\cA_i'\cA_o\cB_i'\cB_o}  X^{T}_{\cA_i'}}\right)=0, \forall \Tr{X}=0,&\\
\text{Tr}_{\cB_o\cB_i'}\left({\Omega_{\cA_i'\cA_o\cB_i'\cB_o}  Y^{T}_{\cB_i'}}\right)=0, \forall \Tr{Y}=0,&
\end{align*}
where $\mathbb{I}$ is the identity matrix of appropriate dimension, $X$ and $Y$ are Hermitian operators,
 and $X^T$ is the transpose of $X$.
\textcolor[rgb]{0.00,0.00,0.00}{ The first and second constraints require $\Pi$ to be completely positive and trace-preserving; the third and fourth constraints mean that $\Pi$ is non-signalling from both Alice to Bob and Bob to Alice.
}


\begin{figure}[ht]
\[  \includegraphics[width=5cm]{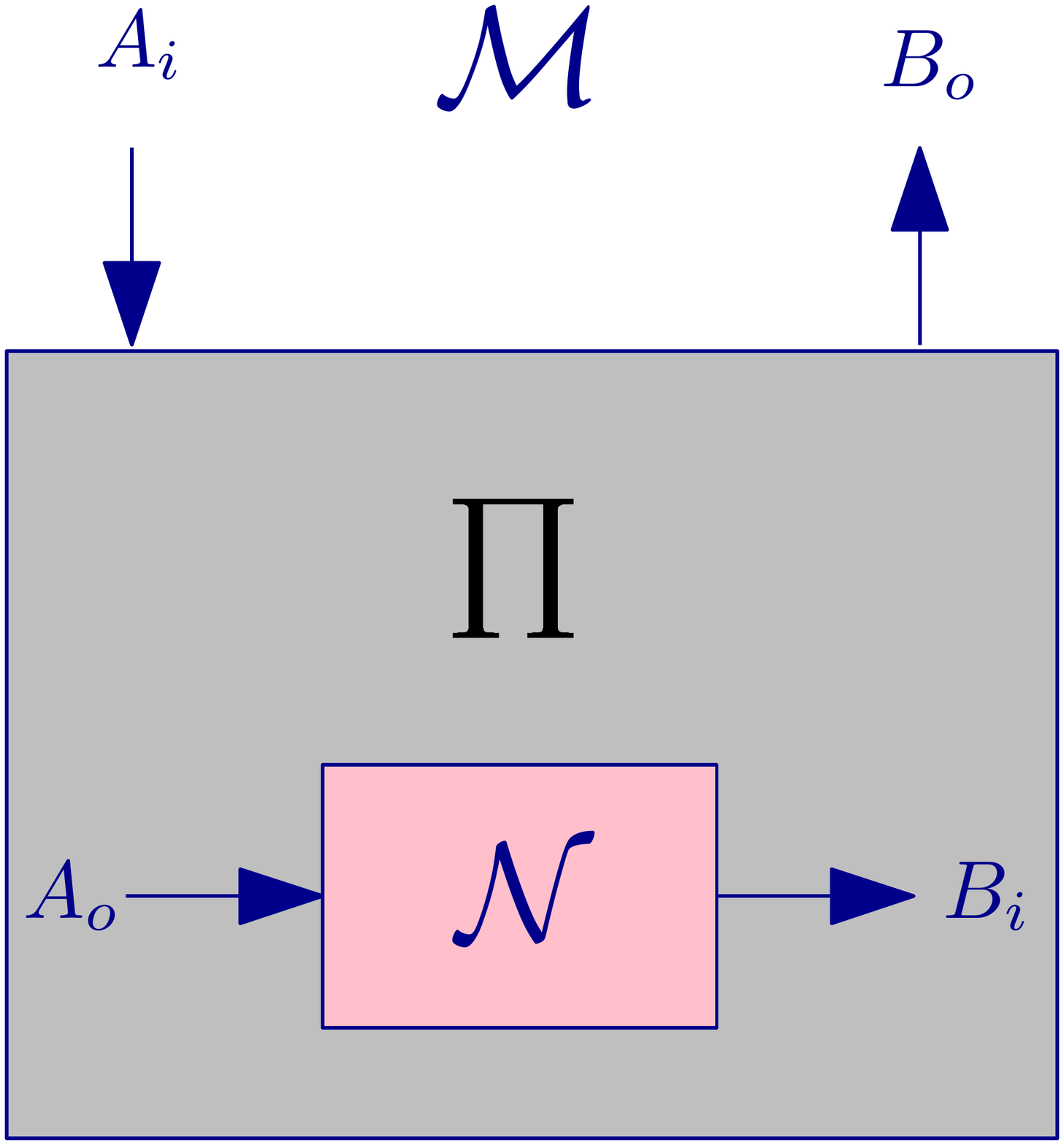}\]
  \caption{A general simulation network:  implementing a channel $\cM$ using another channel $\cN$ once, and the QNSC $\Pi$ between Alice and Bob. 
}
\label{fig:QNSC}
\end{figure}

Suppose $\cN: \ket{k}\bra{k}\rightarrow \rho_k \in \cL(\cB)$ is a CQ channel that maps a set of classical states $\ket{k}\bra{k}$ for $k=0, \dots, n-1$ into some quantum states $\rho_k\in \cL(\cB)$.
Suppose that $P_k$ are the orthogonal projectors onto the support of $\rho_k$, respectively.
Then $\{P_k\}$ defines a confusability graph $G$ with  vertex set $\mathbb{Z}_n$ and
two vertices $i$ and $j$  are connected if and only if $\Tr{ P_iP_j}\neq 0$.

Let $\cM$ be the composition channel of $\cN$ and a QNSC $\Pi$ as illustrated in Fig. \ref{fig:QNSC}.
The \emph{one-shot} zero-error classical capacity of $\cN$ assisted by $\Pi$
is the logarithm of the largest integer $m$ so that $\cM$ can simulate a noiseless classical channel that can send $m$ messages.
In \cite{DW14}, Duan and Winter showed that this one-shot capacity  is $\log \lfloor \Upsilon(\cN)\rfloor$,
where $\Upsilon(\cN)$ is the value of the following SDP with variables $s_k\in \mathbb{R}$ and $R_k\in \cL(\cB)$:
\begin{align}
\Upsilon(\cN)=&\max \sum_k s_k \notag\\
 \text{subject to: } &s_k\geq 0, \notag\\
  &0\leq R_k \leq s_k(\mathbb{I}-P_k), \label{eq:SDP}\\
&\sum_k\left(s_kP_k+R_k\right)=\mathbb{I}. \notag
\end{align}
\textcolor[rgb]{0.00,0.00,0.00}{It is not difficult to see that $\Upsilon$ is super-multiplicative \cite{DW14}:
\begin{align}
\Upsilon(\cN_1\otimes \cN_2)\geq \Upsilon(\cN_1)\Upsilon(\cN_2). \label{eq:Upsilon super multi}
\end{align}}
For an arbitrary graph $G$,
Duan and Winter considered the case of asymptotically many channel uses
and showed that $$\min_{\cN} \lim_{m\rightarrow \infty}\frac{1}{m}\log \Upsilon(\cN^{\otimes m})=\log \vartheta(G),$$
where the minimization is over all CQ channels $\cN$  with confusability graph $G$.

Herein we try to determine $\Upsilon(\cN)$.
Apparently $\Upsilon(\cN)\geq \alpha(G)$, the independence number of $G$. This lower bound can be achieved as follows. We choose a maximum independent set $\cal{I}$ of size $\alpha(G)$
and set $s_k=1$ if $k\in \cal{I}$ and $s_k=0$, otherwise. For some $s_{k^*}=1$, let $R_{k^*}= \mathbb{I}- \sum_{k\in \cal{I}} P_k$ and $R_k=0$ for $k\neq k^*$.
Then the constraints of (\ref{eq:SDP}) are satisfied and $\Upsilon(\cN)\geq \sum_k s_k=\alpha(G).$

To find an upper bound on $\Upsilon(\cN)$, we consider the dual problem of (\ref{eq:SDP}):
\begin{align}
\hat{\Upsilon}(\cN)=&\min \Tr{T} \notag\\
 \text{subject to: } &\Tr{P_k T}-\Tr{(\mathbb{I}-P_k)Q_k}\geq 1, \notag\\
  &Q_k+T\geq 0, \label{eq:dual_SDP}\\
 &Q_k\geq 0,\notag
\end{align}
where $T\in \cL(\cB)$ is Hermitian. It can be verified that \[\Tr{T} -\sum_k\Tr{ R_k(T+Q_k)}\geq \sum_k s_k \]
and the duality gap is zero when $\Tr{ R_k(T+Q_k)}=0$ for $s_k\neq 0$.
By choosing $Q_k=0$ for all $k$  and $T=\eta(\{P_k\})\sigma$, where $\sigma$ is the handle of $\{P_k\}$,  we have $$\hat{\Upsilon}(\cN)\leq \eta(\{P_k\}).$$
When $\{P_k\}$ is an OOR of $G$, we have
\begin{align}
\hat{\Upsilon}(\cN)\leq \vartheta(G). \label{eq:Upsilon upper bound}
\end{align}
\textcolor[rgb]{0.00,0.00,0.00}{Note that (\ref{eq:Upsilon upper bound}) is also implied by Lemma 13 and the proof of Theorem 5 in \cite{DW14}.
}
\textcolor[rgb]{0.00,0.00,0.00}{
The asymptotic QNSC-assisted zero-error classical capacity of $\cN$ is upper bounded by $\log \vartheta(G)$:
\begin{align}
C_{0,\text{NS}}(\cN)=\lim_{m\rightarrow \infty}\frac{1}{m}\log \Upsilon(\cN^{\otimes m})\leq  \lim_{m\rightarrow \infty}\frac{1}{m}\log \vartheta(G^{\boxtimes m})= \log \vartheta(G), \label{eq:C0N upper bound}
\end{align}
where the inequality follows from (\ref{eq:Upsilon upper bound}) and $G^{\boxtimes m}$ is the confusability corresponding to $\cN^{\otimes m}$;
the last equality is because that $\vartheta$ is multiplicative \textcolor[rgb]{0.00,0.00,0.00}{(\ref{eq:theta multiplicative})}.
}

It is suspected that equality may hold in (\ref{eq:C0N upper bound}) for graphs with nontrivial automorphisms. In the rest of this article, we will directly solve the SDP (\ref{eq:SDP}) for the CQ channel
$\cN: \ket{k}\bra{k}\rightarrow \ket{u_k}\bra{u_k}$, 
where $\{ \ket{u_k}\}$ is an OOR for some circulant graph $G$, defined by equal-sized cyclotomic cosets.

\section{Circulant Graphs} \label{sec:circulant_graph}
In this section we first discuss the definition of a circulant graph and its properties, and then derive an orthonormal representation $\{\ket{u_k}\}$ with $\ket{u_k}=U^k\ket{u_0}$, where $U$ is a unitary operator.
Then we show that a circulant graph $G$, defined by equal-sized cyclotomic cosets modulo $n$, will induce a CQ channel $\cN$ so that $\Upsilon(\cN)=\vartheta(G)$.
This is done by explicitly constructing $s_k$ and $R_k$, which lead to a feasible solution to the above SDP with object function $\sum_k s_k=\vartheta(G)$.

\subsection{Orthonormal Representation of Circulant Graphs}

Let $C$ be a subset of $ \mathbb{Z}_n\setminus\{0\}$  so that $-C=C$.
A circulant graph $G=X(Z_n,C)$, defined by the connection set $C$, has an edge set $\{ij: i-j\in C\}$.
Consequently its adjacency matrix $A_G$ has entries $[A_G]_{i,j}=1$ if and only if $i-j\in C$.
(For example, a cycle graph $\cC_n$ is defined by the connection set $C=\{1,n-1\}$.)
Define a unitary matrix
\begin{align}
U=& \diag{1, e^{-2\pi i/n}, \dots,  e^{-2(n-1)\pi i/n}}. \label{eq:unitary operator}
\end{align}
Let $\ket{\mathbf{1}}=(1\ 1\cdots 1)$ be the vector whose entries are all ones of appropriate dimension.
It can be easily verified that the eigenvectors of $A_G$ are
$\ket{v_k} = U^{-k} \ket{\mathbf{1}}$ 
with corresponding eigenvalues
\begin{align}
\lambda_k=\sum_{j\in C} e^{2\pi ijk/n} \label{eq:eigenvalue}
\end{align}
for $k=0, \dots,n-1.$
Let $\lambda_{\max}$ and $\lambda_{\min}$ be the largest and the smallest eigenvalues of $A_G$, respectively.
It is easy to see that $\lambda_{\max}=\lambda_0=|C|$. For a circulant graph $G$ that is edge-transitive, its Lov\'{a}sz number is
$
\vartheta(G)=\frac{-n\lambda_{\min}}{\lambda_{\max}-\lambda_{\min}}
$ \cite{Lov79}.
Note that $\lambda_{\min}<0$ since $\tr{A_G}=0$.
Below we provide an orthonormal representation for an arbitrary circulant graph.

\bt \label{thm:OOR}
Consider a circulant graph $G=X(\mathbb{Z}_n,C)$.
Let $\eta=\frac{-n \lambda_{\min}}{\lambda_{\max}-\lambda_{\min}}$.
Define
\begin{align*}
\ket{u_0}=&\frac{1}{\sqrt{\eta}}\left(1, \sqrt{\frac{\lambda_{1}-\lambda_{\min}}{\lambda_{\max}-\lambda_{\min}}}  ,\dots, \sqrt{\frac{\lambda_{n-1}-\lambda_{\min}}{\lambda_{\max}-\lambda_{\min}}} \right)
\end{align*}
and
\begin{align}
\ket{u_k}= U^k \ket{u_0}, \qquad k=0, \dots, n-1, \label{eq:unitary generating orthonormal}
\end{align}
where 
$U$ is the  unitary operator defined in (\ref{eq:unitary operator}).
Then $\{\ket{u_k}\}$   is an orthonormal representation of the circulant  graph $G$.
Moreover,
\begin{align*}
\braket{u_{k}}{ u_{k+m}} =\frac{[A_G]_{k+m,k}}{-\lambda_{\min}}+\delta_{m,0}
\end{align*}
for any $k$, where $\delta_{m,j}$ is the Kronecker delta function.
\textcolor[rgb]{0.00,0.00,0.00}{If $G$ is edge-transitive, then $\{\ket{u_k}\}$ is an OOR with value $\eta=\vartheta(G)$ and handle $\ket{c}=(1,0,\dots,0)$.
}

\etp
\begin{proof}
It is straightforward to verify that $\{\ket{u_k}\}$ is an orthonormal representation:
\begin{align*}
\braket{u_k}{u_{k+m}} =&\frac{1}{\vartheta(G)} \sum_{j=0}^{n-1} \frac{\lambda_j-\lambda_{\min}}{\lambda_{\max}-\lambda_{\min}} e^{-2\pi ij m /n}\\
=& - \frac{1}{n\lambda_{\min}}\sum_{j=0}^{n-1} \lambda_j e^{-2\pi ij m /n}+\frac{1}{n}\sum_{j=0}^{n-1} e^{-2\pi ij m /n}\\
=& \frac{1}{-\lambda_{\min}} \sum_{j=0}^{n-1}\sum_{l\in C} \frac{e^{2\pi ij (l- m) /n}}{n} +\delta_{m,0}\\
=& \frac{1}{-\lambda_{\min}} \sum_{l\in C}\sum_{j=0}^{n-1} \frac{e^{2\pi ij (l- m) /n}}{n} +\delta_{m,0}\\
=& \frac{[A_G]_{k,k+m}}{-\lambda_{\min}}+\delta_{m,0}.
\end{align*} 
If $G$ is edge-transitive, $\vartheta(G)=\eta$  \cite{Lov79}, $\{\ket{u_k}\}$ is an OOR of $G$ and $\ket{c}=(1,0,\dots,0)$ is the handle:
\[
\frac{1}{| \braket{c}{u_k}|^2} =\vartheta(G), \qquad k=0,\dots, n-1.
\]


\end{proof}
\textbf{Remark:} If $\lambda_{\min}$ is of multiplicity $\mu$, then $\mu$ entries of $\ket{u_k}$ are zeros.
Also, it is straightforward to see that a graph with an orthonormal representation in the form of (\ref{eq:unitary generating orthonormal}) must be circulant.

\subsection{Circulant Graphs defined by Cyclotomic Cosets Modulo $n$}

In the following we will define circulant graphs by cyclotomic cosets modulo $n$.
Cyclotomic cosets usually appear in the application of coding theory for  minimal polynomials over finite fields or integer rings \cite{MS77}.
We use a more general concept here. 

Let $\mathbb{Z}_n^\times=(\mathbb{Z}/n\mathbb{Z})^\times$ denote the multiplicative group of $\mathbb{Z}_n$,
which consists of the units in $\mathbb{Z}_n$ and its size is determined by the  Euler's totient function:  $|\mathbb{Z}_n^\times|= \varphi(n)$.
Suppose $q\in \mathbb{Z}_n^\times$.
The cyclotomic coset modulo $n$ over $q$ which contains $s\in \mathbb{Z}_n$ is
$$C_{(s)=}\{ s, sq, sq^2, \dots, sq^{r_s-1}\},$$
where $r_s$ is the smallest positive integer $r$ so that $sq^{r}\equiv s \mod n$.
The subscript $s$ is called the coset representative of $C_{(s)}$.
Since $q$ and $n$ are relatively prime, we have $q^{\varphi(n)}\equiv 1\mod n$  by Fermat-Euler theorem.
Thus $r_s$ exists for any $s$ and the cyclotomic cosets are well-defined:
$C_{(\alpha)}=C_{(\beta)}$ if and only if $\alpha=\beta q^c \mod n$ for some $c\in \mathbb{Z}$.
 Any element in a coset can be the coset representative, though it is usually  the smallest number in the coset.
As a consequence, the integers modulo $n$ are partitioned into disjointed cyclotomic cosets:
$$
\mathbb{Z}_n= \bigcup_{j=0}^t C_{(\alpha_j)},
$$
where $\{\alpha_0=0, \alpha_1, \dots, \alpha_t\}$ is  a set of (disjointed) coset representatives.
We consider $t> 1$, while the case $t=1$ is trivial.
Since $q$ is relatively prime to $n$, we always have $C_{(0)}=\{0\}$.
It suffices to consider partitions of $\mathbb{Z}_n^*=\mathbb{Z}_n\setminus \{0\}$.

If $-1\in C_{(1)}$, we can define a circulant graph $X(\mathbb{Z}_n, C_{(\alpha)})$ for any $\alpha\neq 0$.
Assume further that these cyclotomic cosets  are \emph{equal-sized}, except $C_{(0)}=\{0\}$.
That is, $|C_{(\alpha)}|=|C_{(1)}|$ for any $\alpha\neq 0$, and $n= t |C_{(1)}|+1$.
A circulant graph defined by these cyclotomic cosets have some interesting properties that are critical to the proof of our main theorem.
First, by (\ref{eq:eigenvalue}), the eigenvalues of $X(\mathbb{Z}_n, C_{(\alpha_j)})$ are
\[
\lambda_{k}^{(\alpha_j)}=\sum_{l\in C_{(k\alpha_j)}}e^{2\pi i l/n},
\]
which depends only on its cyclotomic coset.
Each $\lambda_{k}^{(\alpha_j)}$ is of multiplicity $|C_{(1)}|$, except for $\lambda_0$, which is of multiplicity $1$.
It can be seen that these graphs $X(\mathbb{Z}_n, C_{(\alpha_j)})$ are equivalent and it suffices to consider $G=X(\mathbb{Z}_n, C_{(1)})$.

On the other hand,
suppose $\beta\in \mathbb{Z}_n^{\times} \setminus C_{(1)}$.
Let $\tau_{\beta}(C_{(\alpha)})= C_{(\alpha\beta)}$.
It can be checked that $\tau_{\beta}$ is a permutation on the cyclotomic cosets of order at most $t$.
One can delve into more about the structure of $\tau_{\beta}$, but we only need the following equation in the proof of our main theorem:
\begin{align}
\mathbb{Z}_n= \bigcup_{j=0}^t C_{(\alpha_j)} = \bigcup_{j=0}^t C_{(\alpha_j\beta)}.
\end{align}
(Note that the indices are under modulo $n$ and we will always omit ``\text{mod} n" as it is clear from the context.)
\be

\begin{figure}[!h]
\begin{center}
  \includegraphics[width=7cm]{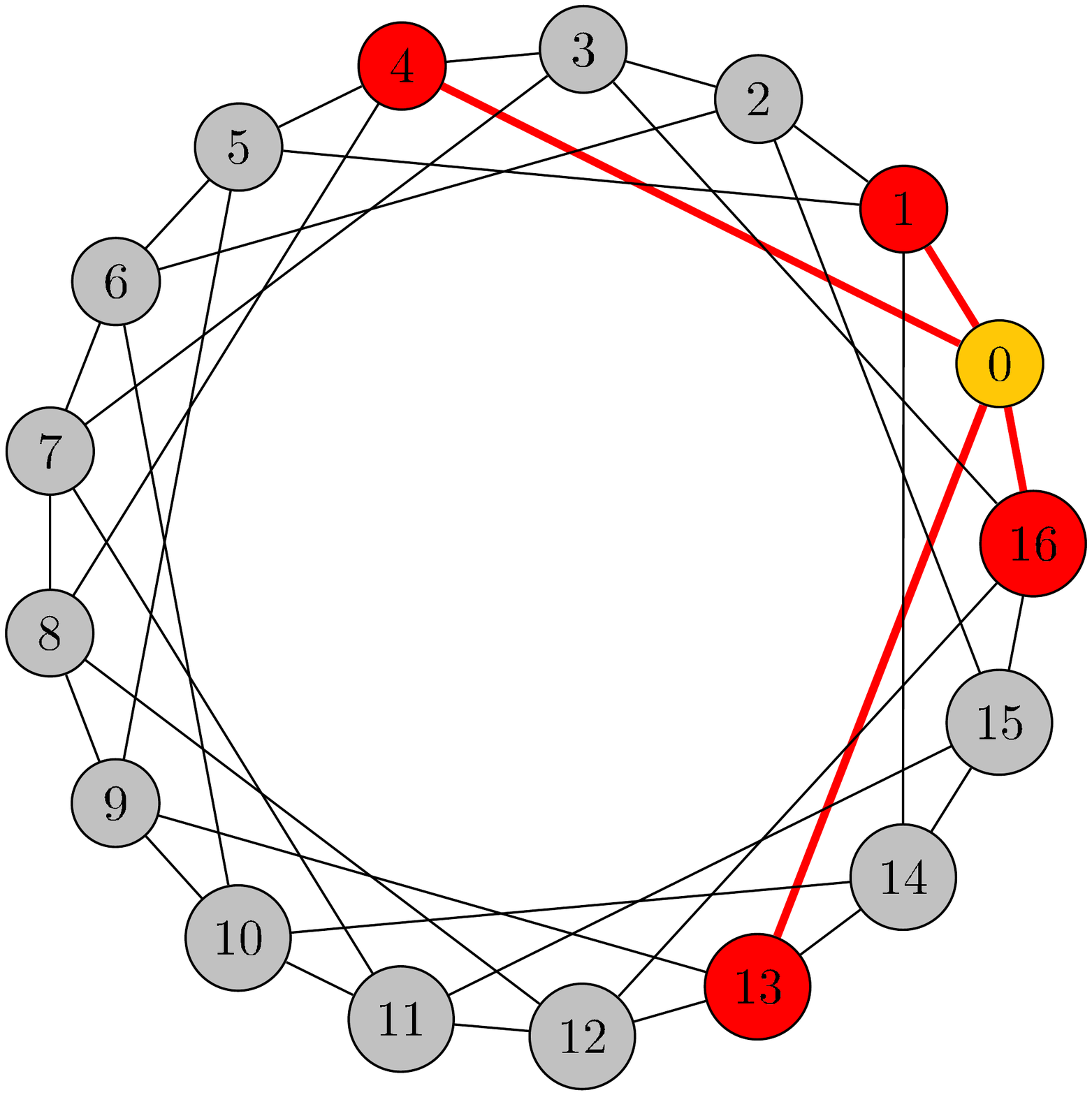}
  \caption{The circulant graph $X(\mathbb{Z}_{17}, \{1,13,16,4\})$). 
}\label{fig:circulant17}
\end{center}
\vspace{-0.4cm}
\end{figure}

For $\mathbb{Z}_{17}^{\times}=\langle 3 \rangle$, $-1\equiv 3^{8}$ and $13\equiv 3^{4}$. Let $C_{(1)}=\langle 13 \rangle$ and we have
\begin{align*}
C_{(0)}=&\{0\},\\
C_{(1)}=&\{1,13,16,4\},\\
C_{(2)}=&\{2,9,15,8\},\\
C_{(3)}=&\{3,5,14,12\},\\
C_{(6)}=&\{6,10,11,7\}.
\end{align*}
The circulant graph $X(\mathbb{Z}_{17}, C_{(1)})$ is shown in Fig. \ref{fig:circulant17}.
\definecolor{reddishyellow}{cmyk}{0,0.22,1.0,0.0}

\eep

Now we are ready to derive our main theorem.
Characterization of equal-sized cyclotomic cosets is left to the next section.

\bt \label{thm:main}
Suppose $\mathbb{Z}_n^*= \bigcup_{j=1}^t C_{(\alpha_j)}$, where $\{C_{(\alpha_j)}\}$ are cyclotomic cosets modulo $n$ over $q$ of equal size for some $q$ relatively prime to $n$ and $C_{(1)}=C_{(-1)}$.
Let $\cN$ be the CQ channel induced by the orthonormal representation $\{\ket{u_k}\}$ of $G=X(\mathbb{Z}_n, C_{(1)})$ in Theorem \ref{thm:OOR}. Assume further that $G$ is edge-transitive.
Then $$ \Upsilon(\cN)=\vartheta(G)$$
and
$$C_{0,\text{NS}}(\cN)=\log \vartheta(G).$$
Moreover, an optimal  solution  to the SDP (\ref{eq:SDP}) is
\begin{align}
s_k=\frac{1}{n}\vartheta(G),\ R_k= U^k R_0 U^{-k},\ R_0=\frac{1}{n}\left(\mathbb{I}-\sum_{j=0}^{n-1}x_j P_j\right), \label{eq:R_k}
\end{align}
where $U$ is defined in (\ref{eq:unitary operator}) and
$x_j=\frac{\lambda_{j\beta}-\lambda_{\beta}}{\lambda_{0}-\lambda_{\beta}},$
given  $\lambda_\beta= \lambda_{\min}$ for some $\beta\in \mathbb{Z}_n^{\times}$.
(In particular, $x_0=1$ and $x_j=0$ for $j\in C_{(1)}$.)


\etp
\begin{proof}
Apparently, $\Upsilon(\cN)=\sum_k s_k=\vartheta(G)$. Also,
\begin{align*}
\sum_k\left(s_kP_k+R_k\right)&=\frac{1}{n} \sum_k \left(\vartheta(G))P_k-\sum_{j=0}^{n-1}x_j U^{k}P_jU^{-k}\right) + \mathbb{I}\\
&=\frac{1}{n} \left(\vartheta(G)-\sum_{j=0}^{n-1}x_j \right) \sum_k P_k + \mathbb{I}\\
&=\frac{1}{n}  \left(-\sum_{j=0}^{n-1} \frac{\lambda_{j\beta}}{\lambda_{0}-\lambda_{\beta}} \right) \sum_k P_k + \mathbb{I}\\
&=\mathbb{I},
\end{align*}
where the last equality is because $\sum_{j=0}^{n-1} \lambda_{j\beta}= \sum_{j=0}^{n-1} \lambda_{j}=0$.
It remains to verify $0\leq R_0\leq s_0(\mathbb{I}-P_0)$.

Let $D=\sum_{j=0}^{n-1}x_j P_j$.
From Theorem \ref{thm:OOR}, we have
$$
[P_j]_{a,b}=\frac{1}{\vartheta(G)}\sqrt{\frac{(\lambda_a-\lambda_{\beta})(\lambda_b-\lambda_{\beta})}{(\lambda_0-\lambda_{\beta})^2}} e^{-2\pi i j(a-b)/n}.
$$
Thus for $a\neq b$,
\begin{align*}
[D]_{a,b}
=&\frac{1}{\vartheta(G)}\sqrt{\frac{(\lambda_a-\lambda_{\beta})(\lambda_b-\lambda_{\beta})}{(\lambda_0-\lambda_{\beta})^2}}  \sum_{j=0}^{n-1}\frac{\lambda_{j\beta}-\lambda_{\beta}}{\lambda_0-\lambda_{\beta}}  e^{-2\pi i j(a-b)/n}\\
=&\frac{1}{-n\lambda_{\beta}}\sqrt{\frac{(\lambda_a-\lambda_{\beta})(\lambda_b-\lambda_{\beta})}{(\lambda_0-\lambda_{\beta})^2}}  \sum_{j=0}^{n-1} \lambda_{j\beta} e^{-2\pi i j(a-b)/n}\\
=&\frac{1}{-n\lambda_{\beta}}\sqrt{\frac{(\lambda_a-\lambda_{\beta})(\lambda_b-\lambda_{\beta})}{(\lambda_0-\lambda_{\beta})^2}}  \sum_{j=0}^{n-1} \sum_{k\in C_{(\beta)}} e^{2\pi i j(k-(a-b))/n}\\
=&\left\{
              \begin{array}{ll}
                \frac{1}{-\lambda_{\beta}}\sqrt{\frac{(\lambda_a-\lambda_{\beta})(\lambda_b-\lambda_{\beta})}{(\lambda_0-\lambda_{\beta})^2}}, & \hbox{ if $a-b\in C_{(\beta)}$;} \\
                0   , & \hbox{if $a-b\neq C_{(\beta)}$.}
              \end{array}
            \right.
\end{align*}
Similarly, we have
\[
[D]_{a,a}= \sqrt{\frac{(\lambda_a-\lambda_{\beta})(\lambda_a-\lambda_{\beta})}{(\lambda_0-\lambda_{\beta})^2}}
\]
for $0\leq a\leq n-1$.
Therefore, $D$ is a nonnegative matrix.
Observe that $\ket{u_0}$ is a positive eigenvector of $D$ with eigenvalue $1$.

\textcolor[rgb]{0.00,0.00,0.00}{\textbf{Claim:}
the largest eigenvalue of $D$ is $1$.}

As a consequent, $R_0= \frac{1}{n}\left(\mathbb{I}- D\right)\geq 0$.
Also, $R_0\leq s_0(\mathbb{I}-P_0)$ as long as $\vartheta(G)\geq 1$.
\textcolor[rgb]{0.00,0.00,0.00}{Therefore, 
$$C_{0,\text{NS}}(\cN)=\lim_{m\rightarrow \infty}\frac{1}{m}\log \Upsilon(\cN^{\otimes m})\geq  \lim_{m\rightarrow \infty}\frac{1}{m}\log \Upsilon^m(\cN)= \log \vartheta(G),$$
where the inequality is because $\Upsilon$ is super-multiplicative (\ref{eq:Upsilon super multi}).}
Combining with (\ref{eq:C0N upper bound}), we have $C_{0,\text{NS}}(\cN)=\log \vartheta(G).$

\textcolor[rgb]{0.00,0.00,0.00}{It remains to prove the claim.
Let $\ket{u_0}=(a_0, a_1, \dots, a_{n-1})$. Then define $V=\diag{a_0, a_1, \dots, a_{n-1}}$, which is an invertible matrix since $a_j\geq 0$.
Define $B=V^{-1}DV$, which has  the same eigenvalues as $D$. Also $$B\ket{\textbf{1}}=V^{-1}DV\ket{1}= V^{-1}D\ket{u_0}=V^{-1}\ket{u_0}=\ket{\bf{1}}.$$
Since $D$ is nonnegative, $B$ is also nonnegative, so every row sum of $B$ is $1$.
As a corollary of Gershgorin's disk theorem, we know that the largest eigenvalue of a nonnegative matrix is upper bounded by its largest row sum.
Thus the largest eigenvalue of $B$ is exactly }$1$.

\end{proof}

\section{Characterization of Equal-sized Cyclotomic Cosets} \label{sec:characterization}
In this section we characterize some properties of the equal-sized cyclotomic cosets. Then we provide three families of graphs that fit Theorem \ref{thm:main}: the cycle graphs, the Paley graphs, and the cubic residue graphs.

Observe that the cyclotomic coset of $1$ modulo $n$ over $q$ is $C_{(1)}=\langle q \rangle$, which is a cyclic subgroup of the multiplicative group $\mathbb{Z}_n^\times$.
Thus $|C_{(1)}| $  divides $\varphi(n)$.
Since $C_{(1)}=C_{(-1)}$, $|C_{(1)}|$ is even, which implies $n$ is odd.
Consequently, $|C_{(1)}|$ is a common divisor of $\varphi(n)$ and $n-1$.
Let $$\Gamma_d^n=\{ a\in \mathbb{Z}^*_{n+1}: \gcd(a,n)=n/d\}$$ and then $|\Gamma_d^n|=\varphi(d)$. We have $\mathbb{Z}_n= \bigcup_{d:d|n} \Gamma_d^n$.
For each $\alpha\neq 0$, $C_{(\alpha)} \subseteq \Gamma_{d_\alpha}^n$ for some $d_\alpha|n$. 
Therefore we have the following lemma.
\bl \label{lemma:equal_size}
If $\{C_{\alpha_1},\dots, C_{\alpha_t}\}$ is a set of equal-sized cyclotomic cosets modulo $n$,
then $|C_{(1)}|$ must be a common divisor of $\varphi(d)$ for all $d|n$ and $d>1$.
\elp
It remains to find conditions so that $C_{(1)}=C_{(-1)}$.
In the following we provide several families of graphs.

\textbf{Remark:} Lemma \ref{lemma:equal_size} is a necessary condition that equal-sized cyclotomic cosets modulo $n$ exist for a certain $n$.
It is likely also a sufficient condition. However,
we did not find composite $n$   so that the nontrivial equal-sized cyclotomic cosets has  $C_{(1)}=C_{{(-1)}}$.

\subsection{Trivial Equal-sized Cyclotomic Cosets}
For any odd $n\geq 3$, there exists a trivial connection set $C_{(1)}=\{1,n-1\}$, which is a cyclotomic coset modulo $n$ over $n-1$.
\be \label{ex:C7}
For $n=7$ and $q=6$,
we have
\begin{align*}
C_{(0)}=&\{0\},\\
C_{(1)}=C_{(6)}=&\{1,6\},\\
C_{(2)}=C_{(5)}=&\{2,5\},\\
C_{(3)}=C_{(4)}=&\{3,4\}.
\end{align*}
Each of the coset, except $C_{(0)}$, defines a circulant graph equivalent to the cycle graph $\cC_7$.

If  $\cN_1: \ket{k}\bra{k}\rightarrow \rho_k \in \cL(\cB)$ is a CQ channel induced from the OOR of $\cC_7$ as in Theorem \ref{thm:OOR}, then $\rho_k$ is a state in a $5$-dimensional Hilbert space
and we have $\Upsilon(\cN_1)=\vartheta(\cC_7)=3.317$.

\eep
As shown in Example \ref{ex:C7}, $C_{(1)}$ defines the cycle graph $\cC_n$ and
we have $\mathbb{Z}_n = \bigcup_{j=0}^{\frac{n-1}{2}}C_{(j)}$. Each nontrivial eigenvalue has multiplicity 2,
as  can be seen from $|C_{(j)}|=2$ for $j\neq 0$,
and
$\lambda_{\min}= \lambda_{\frac{n-1}{2}}=\lambda_{\frac{n+1}{2}}=-2\cos \frac{\pi}{n}$.
\bc Suppose $\cN$ is a CQ channel induced by the OOR of the cycle graph $\cC_n$ as in Theorem \ref{thm:OOR}.  Then
\[\Upsilon(\cN)= \vartheta(\cC_n)=\frac{n \cos \frac{\pi}{n}}{1+\cos\frac{\pi}{n}}.\]
\ecp

\subsection{Nontrivial Equal-sized Cyclotomic Cosets}

When  $n$ is a prime power, $\mathbb{Z}_n^\times$ is cyclic.
Let $\mathbb{Z}_n^\times= \langle \alpha \rangle$, and $\alpha$ is of order $\varphi(n)$.
Consequently, $-1\equiv \alpha^{\varphi(n)/2}$. Therefore, $-1\in C_{(1)}=\langle q \rangle$
if $q= \alpha^b$ for some $b\mid (\varphi(n)/2)$, and then $|C_{(1)}|=\frac{\varphi(n)}{b}$.
It is clear that $\mathbb{Z}_n^\times$ is equally partitioned by $C_{(1)}$.
Furthermore, if $\mathbb{Z}_n^*\setminus \mathbb{Z}_n^\times$ can also  be equally partitioned by $C_{(1)}$, then $X(\mathbb{Z}_n, C_{(1)})$ is defined by equal-sized cyclotomic cosets.

We first consider the case  when $n$ is not a prime.
\bt
Let $n=p^r$ be a prime power.
Suppose $\mathbb{Z}_n^{\times}= \langle \alpha \rangle $ for $\alpha\in\mathbb{Z}_p$.
Then the graph $X(\mathbb{Z}_{p^r}, \langle \alpha^{p^{r-1}}\rangle)$ is defined by equal-sized cyclotomic cosets. 

\etp
\begin{proof}
We have $\varphi(n)=p^{r-1}(p-1)$ and then $\alpha^{p^{r-1}(p-1)}\equiv 1 \mod p$.
Let $C_{(1)}^r$ be the cyclotomic coset modulo $p^r$ over $\alpha^{p^{r-1}}$ that contains $1$.
Thus $|C_{(1)}^r|=p-1$, which divides $\varphi(p^a)$ for $a=1, \dots, r$.
Also, $-1\equiv (\alpha^{p^{r-1}})^{\frac{p-1}{2}} \in C_{(1)}^r$.

Let  $p C= \{ p\alpha: \alpha\in C\}$ for a set $C\subseteq \mathbb{Z}_{p^r}$.
First, we have $\mathbb{Z}_p^{*}=\Gamma_{p}^p = C_{(1)}^1$.  Also $\mathbb{Z}_{p^2}^{*}=\Gamma_{p^2}^{p^2}\cup \Gamma_{p^2}^{p} =\Gamma_{p^2}^{p^2}\cup  p\Gamma_{p^2}^{p}$.
Since $\Gamma^{p^a}_{p^a}= \mathbb{Z}_{p^{a}}^\times$ can be equally partitioned by the cyclotomic coset $C_{(1)}^a$ for any $a$
as in the proof of Theorem \ref{thm:cosets}, $\mathbb{Z}_{p^2}$ can be partitioned into  cosets of size $p-1$.
Observe that
\begin{align}
\mathbb{Z}_{p^{r}}^*=&\Gamma^{p^r}_{p^r}\cup \Gamma^{p^r}_{p^{r-1}}\cup\cdots\cup\Gamma^{p^r}_{p} \notag\\
=&\Gamma^{p^r}_{p^r}\cup p\left\{\Gamma^{p^{r-1}}_{p^{r-1}}\cup\cdots\cup\Gamma^{p^{r-1}}_{p}\right\} \notag\\
=&\Gamma^{p^r}_{p^r}\cup \left\{\bigcup_{j=1}^{r-1} p^{j}\Gamma^{p^{r-j}}_{p^{r-j}}\right\}, \label{eq:cosets decomposition}
\end{align}
where $\Gamma^{p^r}_{p^r}= \mathbb{Z}_{p^{r}}^\times$ can be equally partitioned by the cyclotomic coset $C_{(1)}^r$.
Thus by induction, $\mathbb{Z}_{p^r}^*$ can be partitioned into cosets of size $p-1$.

Let $C_{(1)}^r \mod p^{a}= \{a \mod p^a: a\in C_{(1)}^r\} $. Since $\langle \alpha  \mod p^r \rangle= \mathbb{Z}_{p^r}^\times$,
$\langle \alpha  \mod p^a \rangle= \mathbb{Z}_{p^a}^\times$ for any $a\leq r$.
An interesting property is $$C_{(1)}^r \mod p^{a} = C_{(1)}^a.$$
Therefore, these cosets are exactly the cyclotomic cosets modulo $p^r$ over $\alpha^{p^{r-1}}$ of equal size.
Suppose $\Gamma^{p^{a}}_{p^{a}}$ is partitioned into the cyclotomic cosets $\{C_{(\alpha_1)}^a, C_{(\alpha_{2})}^a, \dots, C_{(\alpha_{p^{a-1}})}^a\}$.
Then by (\ref{eq:cosets decomposition}), the cyclotomic cosets of $\mathbb{Z}_{p^r}^*$ are
$$\{   p^{r-a}C_{(\alpha_j)}^a\}.$$

\end{proof}

\be
For $\mathbb{Z}_{125}=\langle 2 \rangle$, $-1\equiv 2^{50}$ and $57\equiv 2^{25}$. Let $C^3_{(1)}=\langle 57 \rangle$ and we have
\begin{align*}
C^2_{(1)}=&\{1,7,24,18\},\\
C^2_{(2)}=&\{2,14,23,11\},\\
C^2_{(3)}=&\{3,21,22,4\},\\
C^2_{(6)}=&\{6,17,19,8\},\\
C^2_{(9)}=&\{9,13,16,12\}
\end{align*}
and $$C^2_{(5)}=\{5,10,20,15\}= 5\{1,2,4,3\}= 5\{C^1_{(1)}\}.$$
Consequently,
$\mathbb{Z}_{125}^*\setminus \mathbb{Z}_{125}^\times = 5\{ C^2_{(1)}\cup C^2_{(2)}\cup C^2_{(3)}\cup C^2_{(6)}\cup C^2_{(9)}\}\cup 25 C^1_{(1)}$.
\eep

It is simpler for the case that $n$ is a prime.
\bt \label{thm:cosets}
Let $p=2st+1$ be a prime.
Suppose $\mathbb{Z}_p^*= \langle \alpha \rangle $.
Then the graph $X(\mathbb{Z}_p, \langle \alpha^t\rangle)$ is defined by equal-sized cyclotomic cosets. 
\etp
\begin{proof}
In this case $\mathbb{Z}_n^*= \mathbb{Z}_n^\times$ and $\varphi(n)=n-1$.
Since $\alpha^{2st}\equiv 1 \mod p$,
 the cyclotomic cosets modulo $p$  over $\alpha^t$  are $C_{(1)}, C_{(\alpha)},\dots, C_{(\alpha^{t-1})}$.
Also, $-1\equiv (\alpha^{t})^{s} \in C_{(1)}$.
These cosets  are equal-sized  and $\mathbb{Z}_p^*= \bigcup_{j=1}^{t}C_{(\alpha^j)}$.
If $|C_{(\beta)}|<|C_{(1)}|$ for some $\beta$, then $\beta \alpha^{t|C_{(\beta)}|}\equiv \beta$. Since $\beta$ is a unit in$\mathbb{Z}_p^*$,
we must have $\alpha^{t|C_{(\beta)}|}\equiv 1$, which is a contradiction to the order of $\alpha$.
Then the result is straightforward.
\end{proof}
\be
Consider $\mathbb{Z}_{37}=\langle 2\rangle$. The following graphs satisfy the conditions in Theorem \ref{thm:cosets}: $\cC_{37}=X(\mathbb{Z}_{37}, \{1,36\})$, $X(\mathbb{Z}_{37}, \{1,6,36,31\})$,
$X(\mathbb{Z}_{37}, \langle 27\rangle)$,
 $\cC\cR_{37}=X(\mathbb{Z}_{37}, \langle 8\rangle )$, 
$\cQ\cR_{37}=X(\mathbb{Z}_{37}, \langle 4\rangle)$. 
\eep

\subsection{Paley Graphs}

When $t=2$, the cosets in Theorem \ref{thm:cosets} lead to exactly the Paley graphs or the quadratic residue graphs $\cQ\cR_p$.

A nonzero integer $a$ is called a quadratic residue modulo $n$ if $a= b^2 \mod n$ for some integer $b$; otherwise,  $a$ is a quadratic nonresidue modulo $n$.
Note that 0 is neither a quadratic residue, nor a nonresidue.
Suppose p is a prime such that $p\equiv 1 \mod 4$.
Let $Q$  denote the set of quadratic residues modulo $p$ and $N$ the set of nonresidues.
 Since $p\equiv 1 \mod 4$, $-1\in Q$.
Then $\cQ\cR_p=X(\mathbb{Z}_p, Q)$ \cite{GR01}.

Suppose $\alpha$ is a primitive element of $\mathbb{Z}_p$. Then $Q=\{\alpha^c: c \text{ even}\}$ and $N=\{\alpha^c: c \text{ odd}\}$.
It is clear that $|Q|=|N|=(p-1)/2$ and $\mathbb{Z}_p=Q\cup N\cup\{0\}$.
By Eq. (\ref{eq:eigenvalue}) and the formula for quadratic Gauss sum:
$\sqrt{p}=\sum_{j=0}^{p-1}e^{2\pi  i j^2/p}$, 
the eigenvalues of $\cQ\cR_p$ are 
\begin{align*}
\lambda_j
=& \left\{
                                                              \begin{array}{ll}
                                                                (-1+\sqrt{p})/2, & \hbox{if $j\in Q$;} \\
                                                                (-1-\sqrt{p})/2, & \hbox{if $j\in N$;}\\
(p-1)/2,&  \hbox{if $j=0$,}
                                                              \end{array}
                                                            \right.
\end{align*}
The Paley graphs are self-complimentary and consequently  $\Theta(\cQ\cR_p)=\vartheta(\cQ\cR_p)=\sqrt{p}$  \cite[Theorem 12]{Lov79}.
In fact, $\alpha(\cQ\cR_p^{\boxtimes 2})=p$ \cite{MRR78}.  Let  $b\in N$ and then $\{(a, ab \mod p): a\in \mathbb{F}_p\}$ is an independent set of size $p$ in $\cQ\cR_p^{\boxtimes 2}$.
For example, the smallest Paley graph is $\cQ\cR_5= \cC_5$, and $\{(0,0),(1,2),(2,4),(3,1),(4,3)\}$ is an independent set of size five in $\cC_5^{\boxtimes 2}$.
This  shows that the capacity can be achieved by two uses of a channel corresponding to $\cQ\cR_p$.

\bc Suppose $\cN$ is a CQ channel induced by the OOR  of the  Paley graph $\cQ\cR_p$ as in Theorem \ref{thm:OOR}. Then
\[
\Upsilon(\cN) = \vartheta(\cQ\cR_p)=\sqrt{p}.
\]
\ec
\begin{proof}

The proof for Paley graphs is easier than the general proof in Theorem \ref{thm:main} since there are only three cyclotomic cosets and two nontrivial eigenvalues.
The SDP (\ref{eq:SDP}) can be achieved by
 $$ R_0= \frac{1}{p}\left(\mathbb{I}-P_0- \frac{2}{\sqrt{p}+1}\sum_{j\in N}P_j\right).$$
One can show that
\begin{align*}
p[R_0]_{a,b}=\left\{
               \begin{array}{ll}
                 1, & \hbox{if $a=b\in N$;} \\
                 \frac{-1+\sqrt{p}}{1+\sqrt{p}}, & \hbox{if $a=b\in Q$;} \\
                 -\left(\frac{2}{1+\sqrt{p}}\right)^2, & \hbox{if $a,b\in Q$ and $a-b\in N$;} \\
                 0, & \hbox{otherwise.}
               \end{array}
             \right.
\end{align*}
A key observation here is  that
$$
\sum_{b: b\neq a} |p [R_0]_{a,b}| =\left\{
                                  \begin{array}{ll}
                                     p[R_0]_{a,a}, & \hbox{if $a\in Q$;} \\
                                    0, & \hbox{if $a\in N$.}
                                  \end{array}
                                \right.
$$
Then by Gershgorin's disk theorem, the eigenvalues of $pR_0$ are either 1 or
lie in the disks with center $p[R_0]_{a,a}$ and radius $p[R_0]_{a,a}$ for $a\in Q$.
Also note that $R_0$ is Hermitian and it has real eigenvalues.
As a consequence, the eigenvalues of $R_0$ are nonnegative and thus $R_0\geq 0$.

The null space of $R_0$ are spanned by $\sum_{j: j \in N}{\ket{u_j}}$ and $\ket{u_0}$,
which implies $(1,0 ,\dots, 0)$  is an eigenvector of $R_0$ with eigenvalue 0. 

\end{proof}

\subsection{Cubic Residue Graphs}
When $t=3$, the cosets in Theorem \ref{thm:cosets} lead to  the cubic residue graphs $\cC\cR_p$\cite{vanDam96}.
 A nonzero integer $a$ is called a cubic residue modulo $p$ if $a= b^3 \mod p$ for some integer $b$.
The cyclotomic coset $C_{(1)}$ consists of cubic residues.

$\cC\cR_p=X(\mathbb{Z}_p, C_{(1)})$ has three nontrivial eigenvalues, which can be found by the formula for cubic Gauss sum.
These three eigenvalues are the roots of $x^3-3px-ap=0$, where $4p=a^2+b^2$ for some  integers $a\equiv 1\mod 3$ and $b$ \cite{IR90}.
Currently the closed form for $\vartheta(\cC\cR_p)$ is still unknown, since it is related to the determination of Gauss sums \cite{Ito14,BE81}.

These discussions can be extended to $t\geq 4$.

\section{Discussion}\label{sec:discussion}
We have shown that $\Upsilon(\cN)=\vartheta(G)$ for  $\cN$ induced by an OOR $\{\ket{u_k}\}$ of a class of edge-transitive circulant graphs that are defined by equal-sized cyclotomic cosets.
These circulant graphs bear very strong symmetries.
It is interesting to see if there are other graphs that have this property. 
\textcolor[rgb]{0.0,0.00,0.00}{ For graphs with $\vartheta(G)=\alpha(G)$}\footnote{\textcolor[rgb]{0.00,0.00,0.00}{We tried computer search on random graphs and found that $\vartheta(G)=\alpha(G)$ for several asymmetric graphs. It is unknown whether most graphs would have $\vartheta(G)=\alpha(G)$ or $\vartheta(G)>\alpha(G)$. }}, they naturally lead to CQ channels with $\Upsilon(\cN)=\vartheta(G)$.
Now we consider graphs with $\vartheta(G)>\alpha(G)$.

\begin{figure}[!h]
\[
\hspace{-0.8cm}
\begin{array}{cc}
  \includegraphics[width=5cm]{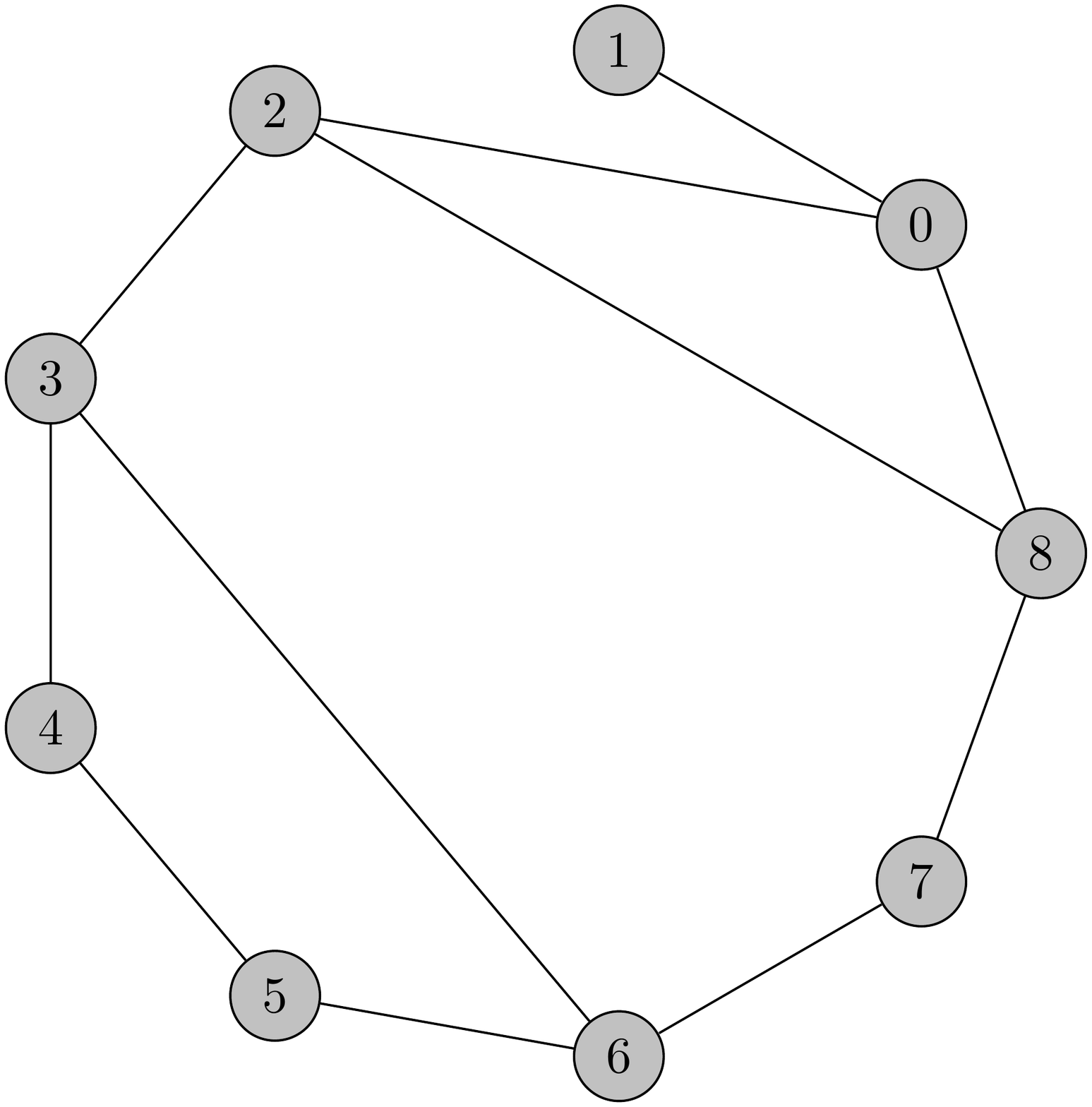}&  \includegraphics[width=5cm]{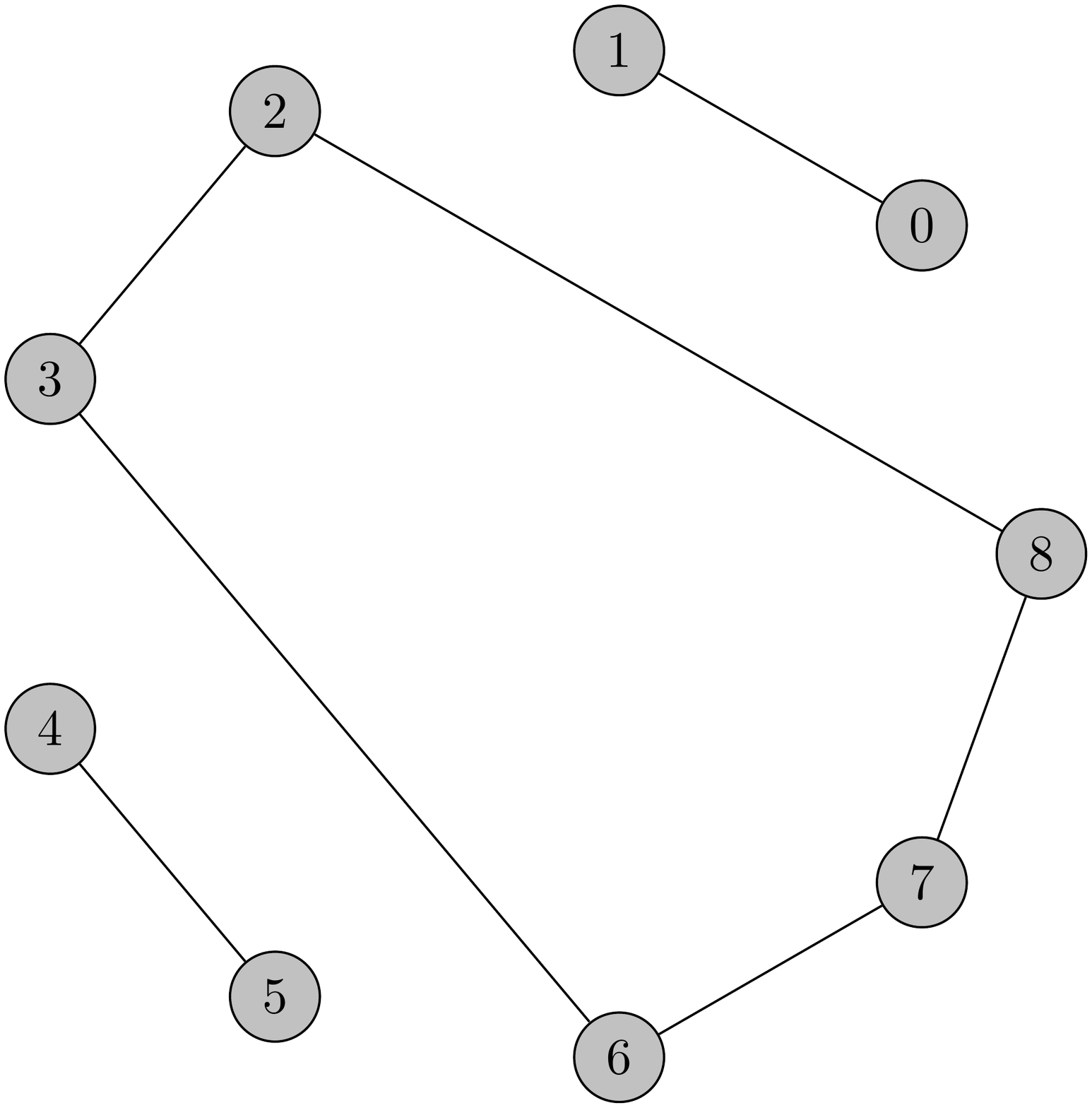}\\
\centering $(a)$ & \centering $(b)$\\
\end{array}
\]
  \caption{(a) an asymmetric graph $G_1$; (b) a degenerate graph ${G}_2$ of $G_1$.}\label{fig:graph9}
\end{figure}

\textcolor[rgb]{0.00,0.00,0.00}{Recall from Definition \ref{def:OR2}, an orthonormal representation of a graph indicates that two vertices are not connected if the trace inner product of their representations is zero. } We say  a graph $G'$ is a \emph{degenerate} graph of $G$ if an orthonormal representation of $G$  is also an orthonormal representation of $G'$, and hence  their Lov\'{a}sz numbers are equal: $\vartheta(G)=\vartheta(G')$. Consequently, if an edge $E\notin G$, $E\notin G'$.
We say a graph $\hat{G}$ is \emph{essential} if it has no proper subgraph $H\subset \hat{G}$ with $\vartheta(H)=\vartheta(\hat{G})$.
Suppose $\{P_k\}$ is an orthonormal representation of the essential graph $\hat{G}$. Then two vertices $i$ and $j$ are connected if and only if $\Tr{P_iP_j}\neq 0$.
\textcolor[rgb]{0.00,0.00,0.00}{Apparently, for any graph $G$, it has an essential subgraph $\hat{G}$  that
 is a subgraph of all degenerate graphs of $G$ by the definition of  orthonormal representation.}

\be \label{ex:degenerate_graph}
Consider Fig. \ref{fig:graph9}, where $G_1$ is an asymmetric graph and $G_2$ is a degenerate graph of ${G}_1$.
The  Lov\'{a}sz numbers are $\vartheta(G_1)=\vartheta({G}_2)=\sqrt{5}+2$. 
Their essential graph is the union of $\cC_5$ and two isolated points: removing any  edge will increase the Lov\'{a}sz number of this graph.
 One can check that $\vartheta(G_1)=\vartheta({G}_2)=\vartheta(\cC_5)+2=\sqrt{5}+2$.
\ee

\begin{figure}[!h]
\[
  \includegraphics[width=4cm]{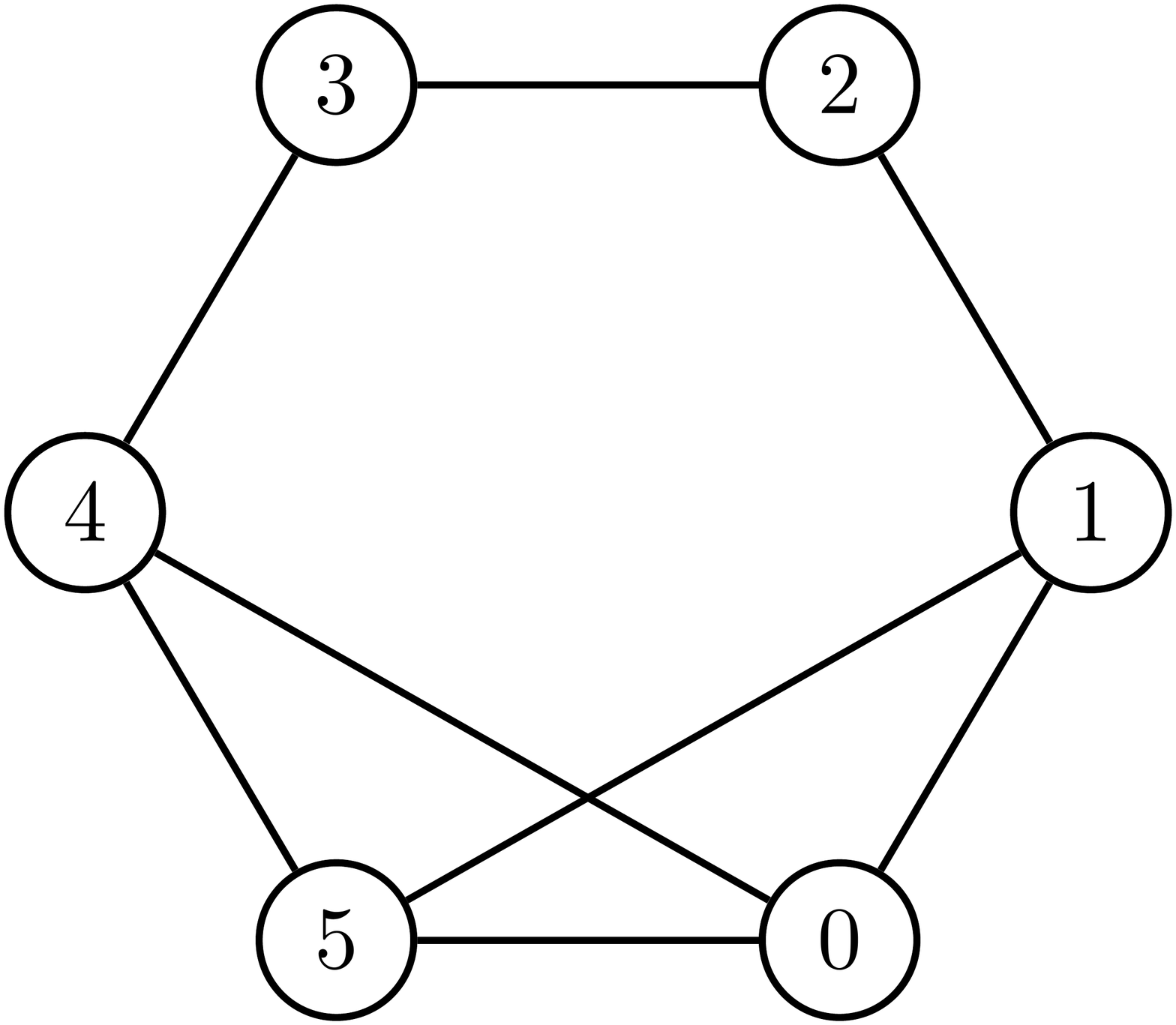}
\]
  \caption{A graph with essential graph $\cC_5$.
}

\label{fig:C5_graph}
\end{figure}

As shown in Example \ref{ex:degenerate_graph}, it is possible that an asymmetric graph has a degenerate graph with nonidentity automorphisms and $\vartheta(G)>\alpha(G)$.
\bc
Suppose a graph $G$ leads to a CQ channel $\cN$ with $\Upsilon(\cN)=\vartheta(G)$.
Then so do the degenerate graphs of $G$.
\ec
Suppose a graph $G$ has an essential graph $\hat{G}$ with $\Upsilon(\hat{\cN})=\vartheta(\hat{G})$.
It is also possible that $G$ leads to  $\Upsilon(\cN)=\vartheta({G})$ as shown in the following example.
\be
Fig. \ref{fig:C5_graph} is a graph $G$ whose essential graph is $\cC_5$.
An orthonormal representation of $\cC_5$ can be easily extended to an orthonormal representation of $G$ by choosing $\ket{u_5}=\ket{u_0}$.
\ee
However, generally it is nontrivial to construct an orthonormal representation of a graph from an orthonormal representation of its proper essential graph.

\begin{figure}[!h]
\[
  \includegraphics[width=4.5cm]{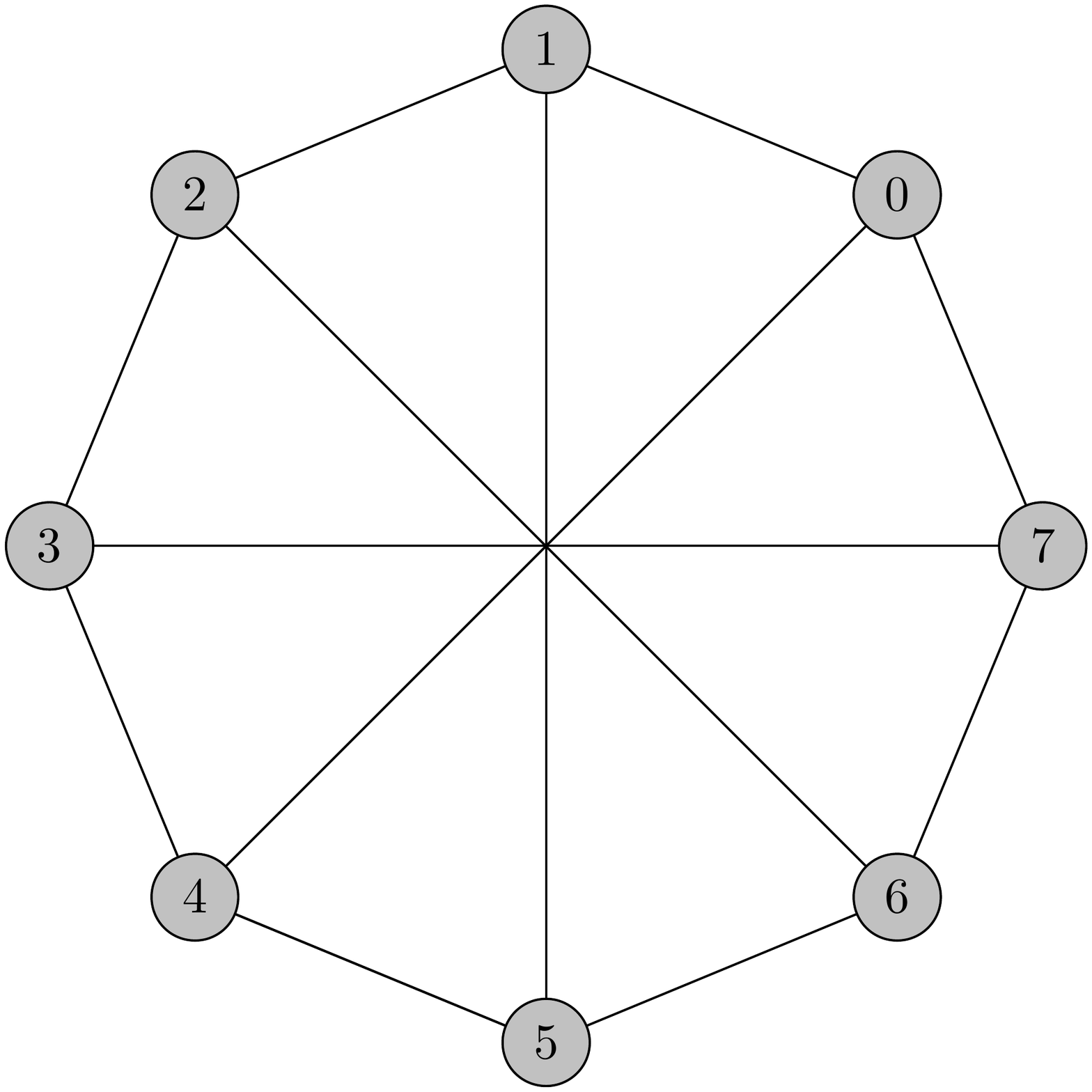}
\]
\vspace{-0.5cm}
  \caption{M\"{o}bius ladder $M_8=X(\mathbb{Z}_8,\{1,4\})$.
}

\label{fig:M8_graph}
\end{figure}

So far we have shown that circulant graphs  defined by equal-sized cyclotomic cosets
and their degenerate graphs would induce CQ channels with $\Upsilon(\cN)=\vartheta(G)$.
Now let consider graphs other than these.
First we show how to find an OOR for any graph $G$, following the proof of \cite[Theorem 3]{Lov79}.
After solving the SDP for $\vartheta(G)$ \cite[Theorem 3]{Lov79},
we end up with  a symmetric matrix $A$ satisfying (\ref{eq:lovasz_matrix}) with the largest eigenvalue $\vartheta(G)$.  
Then  there exist vectors $\ket{x_1}, \dots, \ket{x_n}\in \mathbb{R}^{d+1}$, where $d=\rank{\vartheta(G) I- A}$, such that
\[
\vartheta(G) \delta_{i,j}- [A]_{i,j}=  \braket{x_i}{x_j}
\]
and
the first  entry of $\ket{x_k}$ is 0 for all $k$.
Let $\ket{c}= (1, 0,\dots,0)\in \mathbb{R}^{d+1}$ and
\begin{align}\ket{u_k}= \frac{1}{\sqrt{\vartheta(G)}}(\ket{c}+\ket{x_k}). \label{eq:general OOR}
\end{align}
Then $\{\ket{u_k}\}$
is an OOR of $G$ with value
\[
\vartheta(G)=\frac{1}{|  \braket{c}{u_k}|^2}, \ \forall k.
\]
For a CQ channel induced by $\{\ket{u_k}\}$ to have $\Upsilon(\cN)=\vartheta(G)$ in the SDP (\ref{eq:SDP}), we must have
\begin{align}\bra{c}R_k\ket{c}=0,\ \forall k. \label{eq:necessary condition}
\end{align}
That is, the first row and the first column of $R_k$ are all zeros.
\be \label{ex:M8}
Consider the M\"{o}bius ladder $M_8=X(\mathbb{Z}_8,\{1,4\})$ as shown in Fig. \ref{fig:M8_graph}, which is circulant but beyond the scope of Theorem \ref{thm:main}.
Clearly we may choose $s_k=\vartheta(G)/n$ for all $k$. Let $$R_0= \frac{s_0}{\vartheta(G)}\left( \mathbb{I}- P_0 -\sum_{k=1}^7 x_k P_k\right)$$
and $R_k$ can be defined by permuting the indices of $P_k$ in $R_0$ appropriately.
Since vertices $1, 8,$ and $4$ are neighbors of vertex $0$, we may choose $x_1=x_7=x_4=0$. We define a map: $$\Gamma: i\mapsto n-i.$$
Apparently, $\Gamma$ is an automorphism of $M_8$. Assume $x_2=x_6$ and $x_3=x_5$. By solving the linear system from (\ref{eq:necessary condition}), we have $x_2=x_6=0.5$ and
$x_3=x_5=\frac{\vartheta(G)}{2}-1.$ Surprisingly, $\sum_k x_k= \vartheta(G)$,  $0\leq R_k\leq s_k(\mathbb{I}-P_k)$,  and $\sum_k s_kP_k+ R_k= \mathbb{I}$.
Thus $\Upsilon(\cN)=\vartheta(G)$ for $G=M_8$.

\ee

\begin{figure}[!h]
\[
  \includegraphics[width=4cm]{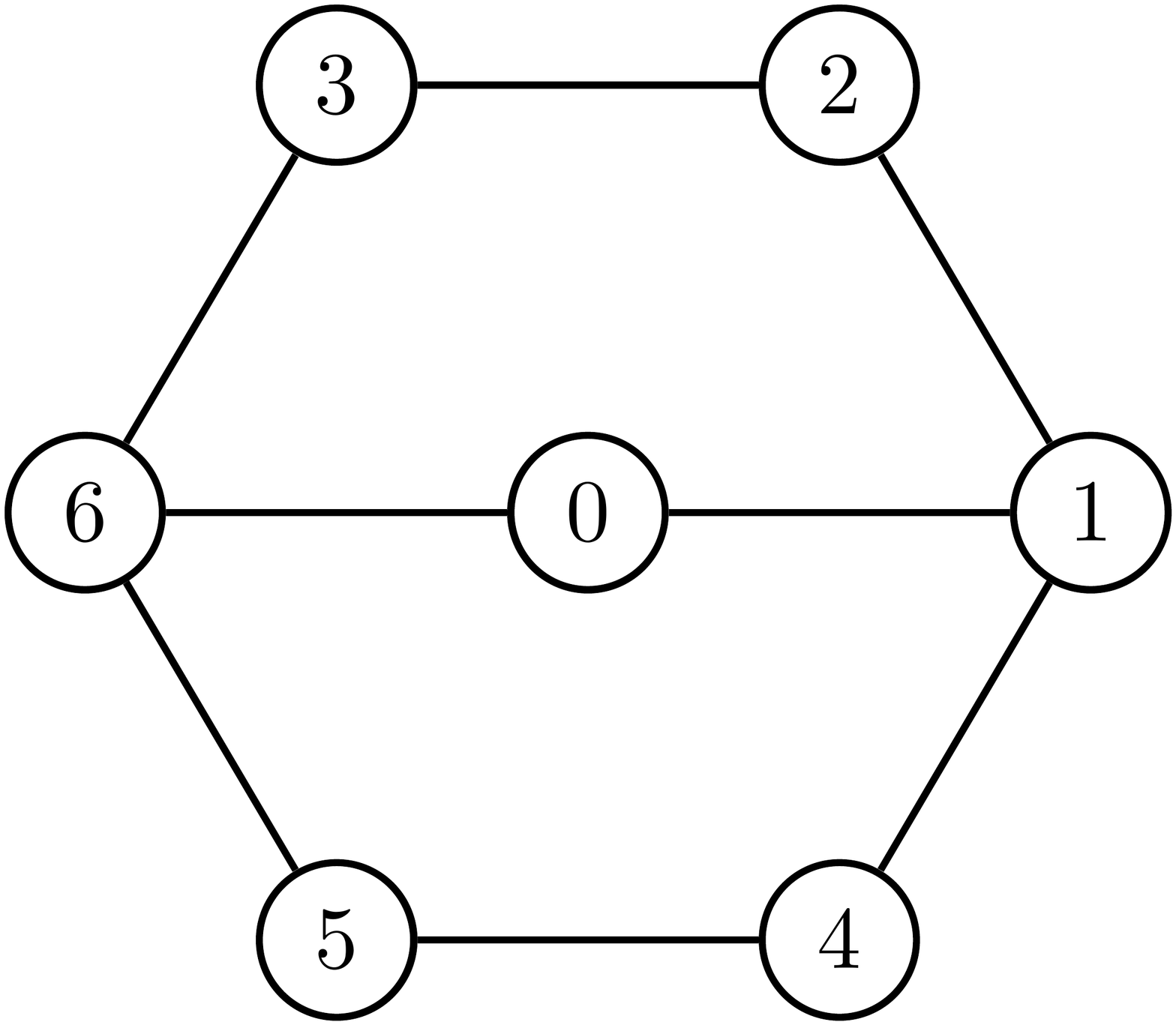}
\]
  \caption{A graph $Z_7$ that is not circulant, regular, or edge-transitive.
}

\label{fig:G7_graph}
\end{figure}

To sum up, we have found many more graphs with  $\Upsilon(\cN)=\vartheta(G)$.
If this holds for general graphs, it would imply that $\lfloor \vartheta(G)\rfloor$ can be achieved by a single channel use\footnote{\textcolor[rgb]{0.00,0.00,0.00}{More precisely, the one-shot capacity is given by $\log \lfloor \Upsilon(\cN)\rfloor$ so the operational meaning of $\Upsilon(\cN)=\vartheta(G)$ may not be clear since the  fractional part of $\vartheta(G)$ will be lost in the one-shot capacity. However, we can always choose a noiseless  $m$-bit channel $\cI_m$ with $\Upsilon(\cI_m)=m$ for $m=10^d$ for some $d$. Then $\cN\otimes \cI_m$ is a channel with one-shot capacity $\log\lfloor \Upsilon(\cN\otimes \cI_m)\rfloor= \log \lfloor m\vartheta(G)\rfloor$,
which is clearly different from $\log\left( m \lfloor \vartheta(G)\rfloor\right)$.}}
In fact, the techniques used in Example \ref{ex:M8} can be generalized to other graphs.
However, we do not know how to prove $R_k\geq 0$.
For example, we have  $\Upsilon(\cN)=\vartheta(G)$ for the graph $G=Z_7$ in Fig. \ref{fig:G7_graph} and its OOR constructed in (\ref{eq:general OOR}).
At the same time, the dual program (\ref{eq:dual_SDP}) is satisfied with $T=\vartheta(G)\ket{c}\bra{c}$ and $Q_k=0$.
Thus there should be a more unifying theory than Theorem \ref{thm:main}
and this is our future research direction.

%

\section*{Acknowledgment}
The authors would like to thank Min-Hsiu Hsieh, Cheng Guo, and Andreas Winter for helpful discussion.
CYL and RD were supported by the Australian Research Council (ARC) under Grant DP120103776.
RD was also supported by the  ARC Future Fellowship under Grant FT120100449 and the National Natural Science Foundation of China under Grant 61179030.


%

\end{document}